\documentclass[conference]{IEEEtran}
\usepackage{cite}
\usepackage{mciteplus}
\usepackage{amsthm}
\usepackage{graphicx}
\usepackage{float}
\usepackage{epsfig}
\usepackage{psfrag}
\usepackage{epstopdf}
\usepackage{subfigure}
\usepackage{amsmath}
\usepackage{url}
\usepackage{stfloats}
\usepackage{amsmath}
\usepackage{amsfonts}
\usepackage{amssymb}
\usepackage{eqnarray}
\usepackage{subeqnarray}
\usepackage{array}
\usepackage{flushend}
\usepackage{paralist}
\usepackage{suffix}
\usepackage{mathtools}
\usepackage{authblk}
\usepackage{epsfig}
\usepackage{psfrag}
\usepackage{epstopdf}
\usepackage{multirow}
\usepackage{tabularx}

\newenvironment{nospaceflalign*}
 {\setlength{\abovedisplayskip}{0pt}\setlength{\belowdisplayskip}{0pt}%
  \csname flalign*\endcsname}
 {\csname endflalign*\endcsname\ignorespacesafterend}

\begin{document}
\title{On Optimizing the Secrecy Performance of RIS-Assisted Cooperative Networks}
\author[]{Abdullateef Almohamad}
\author[]{Ayman Al-Kababji}
\author[]{Anas Tahir}
\author[]{Tamer Khattab}
\author[]{Mazen Hasna}
\affil[]{Department of Electrical Engineering, Qatar University, Doha, Qatar}
\affil[]{ \normalfont Emails: abdullateef@ieee.org, \{aa1405810, a.tahir, tkhattab, hasna\}@qu.edu.qa}
\maketitle
\begin{abstract}
Employing reconfigurable intelligent surfaces (RIS) is emerging as a game-changer candidate, thanks to their unique capabilities in improving the power efficiency and supporting the ubiquity of future wireless communication systems. Conventionally, a wireless network design has been limited to the communicating end points, i.e., the transmitter and the receiver. In general, we take advantage of the imposed channel state knowledge to manipulate the transmitted signal and to improve the detection quality at the receiver. With the aid of RISs, and to some extent, the propagation channel has become a part of the design problem. In this paper, we consider a single-input single-output cooperative network and investigate the effect of using RISs in enhancing the physical layer security of the system. Specifically, we formulate an optimization problem to study the effectiveness of the RIS in improving the system secrecy by introducing a weighted variant of the secrecy capacity definition. Numerical simulations are provided to show the design trade-offs and to present the superiority of RIS-assisted networks over the conventional ones in terms of the system's secrecy performance.

\end{abstract}
\section{Introduction}\label{sec1}
Throughout the consecutive generations of wireless communication networks, channel response manipulation was nothing but a challenge. Obtaining information about the imposed channel at the transmitter has been employed to smartly precode the transmitted signal in such a way to improve the received signal power, as in massive multiple-input multiple-output (mMIMO) systems. However, having a massive number of antennas and RF chains in mMIMO systems results in a significant high power consumption \cite{mendez2016hybrid}.

Recently, reconfigurable intelligent surfaces (RIS)-assisted networks have been proposed as a promising power efficient solution to support mMIMO systems \cite{huang2019reconfigurable}. Basically, an RIS is a large array of passive reflecting elements capable of intelligently reflecting the impinging signals to achieve a certain objective, such as enhancing the rate, the signal-to-noise ratio (SNR) and/or the coverage probability. Each reflecting element introduces a phase shift that can be controlled independently to passively tune the wireless reflected signal. It should be noted that RISs are different from what is known as large intelligent surfaces (LIS) \cite{hu2018beyond}, which actively transmit power. Noting the large scale of RISs, they are usually deployed on walls, building facades and roofs.

RIS-assisted networks have been investigated in many different settings such as capacity and rate improvement analysis \cite{hu2018capacity,jung2020uplink}, power efficiency optimization \cite{huang2019reconfigurable,fu2019intelligent} and communication reliability \cite{jung2019reliability,chafii2016necessary}. Few recent works have studied the RIS-assisted communications from a physical layer security (PLS) point of view \cite{Dong2020,yang2020deep,9086467}. In general, there are two main research directions under the PLS concept, namely, information-theoretic secrecy and covert communications. While the former focuses on improving the secrecy rate between legitimate and eavesdropping users, the latter considers hiding the existence of communication from being detected by an enemy \cite{lu2019intelligent}. It is worth mentioning here that this paper falls under the former concept, i.e. enhancing the secrecy rate. 

In \cite{Dong2020,yang2020deep,9086467}, an RIS-assisted MIMO system model is considered. The secrecy rate of the system is maximized by optimizing the RIS phase shifting matrix and the beamforming at the base station. The authors in \cite{Dong2020} employed a sub-optimal alternating optimization (AO) method to alternately optimize the RIS phase shifts and the beamforming vector assuming perfect channel state information (CSI) knowledge about all involved channels at the RIS, and the absence of direct links with the legitimate user and the eavesdropper. A similar system is considered in \cite{yang2020deep}, but with noting the high computation complexity of the aforementioned optimization problem, a deep learning algorithm is developed to solve the system secrecy rate maximization problem. A more realistic assumption is made in \cite{9086467}, where the phase shifts at the RIS elements are assumed to be discrete, which further complicates the optimization problem and AO-based methods are adopted. The secrecy rate of a SISO model was investigated in \cite{makarfi2019physical}, which assumes a single-antenna for Alice, Bob and Eve. An analytical solution to the secrecy capacity is provided in a vehicular ad hoc network (VANET), where the authors propose two setups to investigate the PLS, but neither setup accounts for the direct links. 

In this paper, we consider a single legitimate user (Bob) and a single eavesdropper (Eve) RIS-assisted SISO system. The secrecy capacity is investigated in the presence of the direct links between the transmitter and both the legitimate user and the eavesdropper. Specifically, assuming the knowledge of the CSI of all involved channels at the RIS, we analytically derive the maximum achievable secrecy rate by tuning the RIS induced phase shifts. The main contributions of this work can be summarised as follows:
\begin{itemize}
    \item Formulation of a generic multi-objective alternating optimization problem for the secrecy capacity, where a coefficient is introduced to trade-off between improving Bob's channel and degrading Eve's channel.
    \item Finding the analytical solution for the optimal RIS phase shifts.
    \item Finding the optimal eavesdropper channel capacity coefficient as a function of the relative channel qualities in the system (studied via simulations).
\end{itemize}

The reminder of this paper is organized as follows, the considered system model and the applied methodology are presented in Section \ref{sec3}. The analytical results are verified and discussed in Section \ref{sec4}. Finally, concluding remarks, limitations and future research are drawn in \ref{sec5}.

\section{System Model}\label{sec3}
\begin{figure}[!ht]
\centering
\includegraphics[width=0.95\linewidth, trim = {2in 0.5in 2in 1in}, clip]{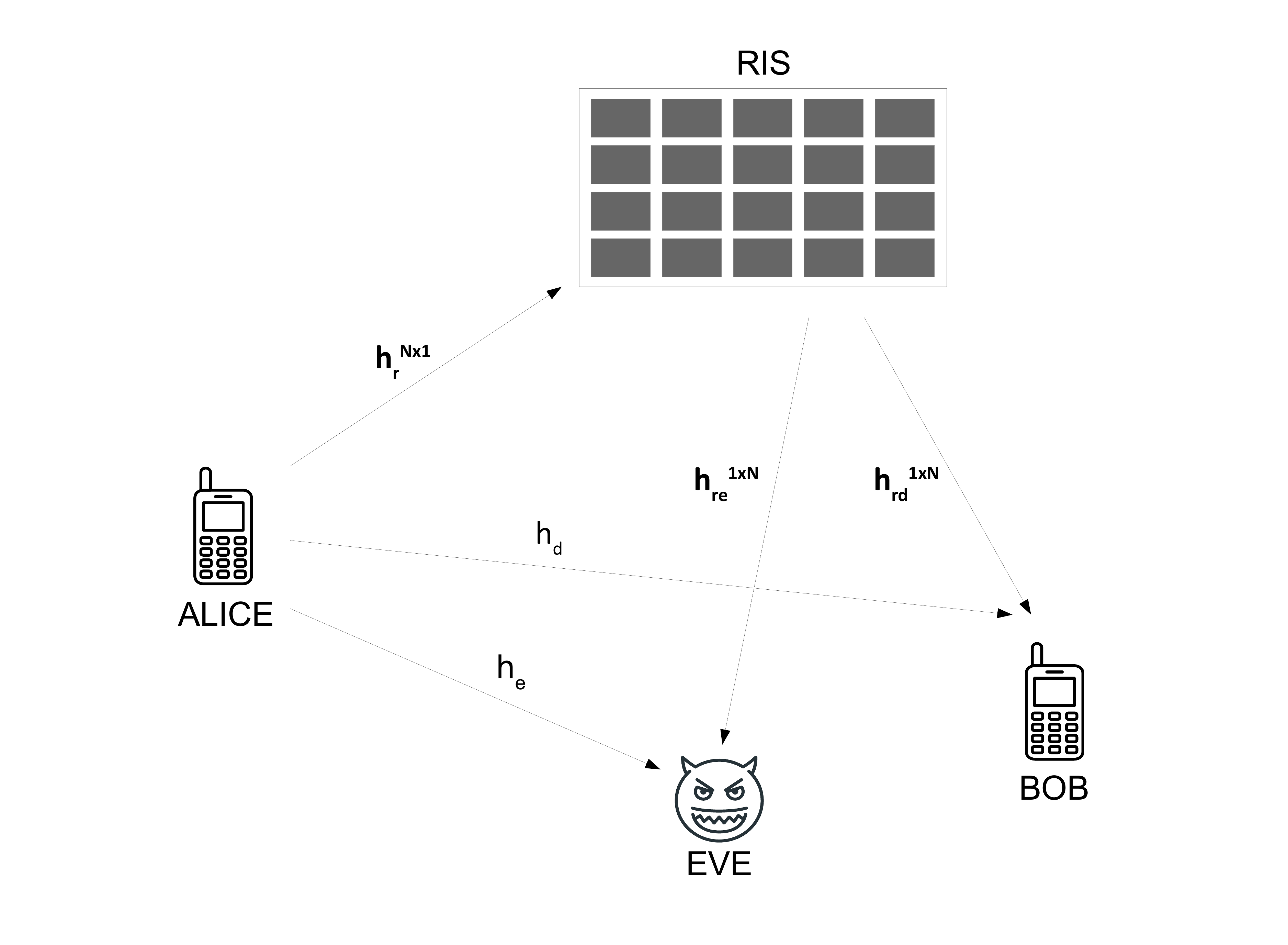} \\
\caption{SISO communication system with a single eavesdropper}
\label{fig:system-model}
\end{figure}

The considered system model consists of a single antenna transmitter, Alice, communicating confidential messages to a single antenna legitimate user, Bob, in the presence of a single antenna eavesdropper, Eve, as shown in Fig.~\ref{fig:system-model}. An RIS with $N$ reflecting elements is installed on a facade of a building in which it reflects the signals between Alice and Bob in such a way to support the system's security. The channels Alice-RIS, Alice-Bob, Alice-Eve, RIS-Bob and RIS-Eve are denoted as ${\bf{h}}_r \in \mathcal{C}^{N\times 1}$, $h_d$, $h_e$, ${\bf{h}}_{rd}\in \mathcal{C}^{1\times N}$ and ${\bf{h}}_{re}\in \mathcal{C}^{1\times N}$, respectively.

The transmitted signal from Alice can be represented as follows
\begin{equation}\label{powerConstraint}
    x=\sqrt{P_t}s,
\end{equation}
where $p_t$ denotes the transmit power and $s$ is a complex symbol drawn from a unit power constellation. Thus, the received signal at Bob is written as
\begin{equation}
    y_d=({\bf{h}}_{rd}{\bf{Q}}{\bf{h}}_{r}+h_d)x+n,
\end{equation}
where $n\sim \mathcal{CN}(0,N_o)$ represents the additive white Gaussian noise, and ${\bf{Q}}=\text{diag}[e^{j\phi_1},\cdots,e^{j\phi_N}]$ where $\phi_i$ denotes the induced phase by the RIS's $i^{\textrm{th}}$ reflecting element. Note that the reflecting elements are assumed to be passive and hence they only affect the phases of the impinging signals. Furthermore, it is assumed that all the channels are independent quasi-static flat-fading with ${\bf{h}}_r=[h_{r1},\cdots,h_{rN}]^T$ and ${\bf{h}}_{rd}=[h_{rd1},\cdots,h_{rdN}]$. Similarly, we can write the received signal at the eavesdropper as follows
\begin{equation}
    y_e=({\bf{h}}_{re}{\bf{Q}}{\bf{h}}_{r}+h_e)x+n,
\end{equation}
with ${\bf{h}}_{re}=[h_{re1},\cdots,h_{reN}]$. The different channels can be written as
\begin{equation}\notag
    {\bf{h}}_{r}: h_{ri} = d_{br}^{-\chi/2}\alpha_ie^{-j\theta_i}
\end{equation}
\begin{equation}\notag
    {\bf{h}}_{rd}: h_{rdi} = d_{rd}^{-\chi/2}\beta_{di}e^{-j\psi_{di}}
\end{equation}
\begin{equation}\notag
    {\bf{h}}_{re}: h_{rei} = d_{re}^{-\chi/2}\beta_{ei}e^{-j\psi_{ei}}
\end{equation}
\begin{equation}\notag
    h_{d}= d_{bd}^{-\chi/2}\eta_{d}e^{-j\tau_{d}}
\end{equation}
\begin{equation}\notag
    h_{e}= d_{be}^{-\chi/2}\eta_{e}e^{-j\tau_{e}}
\end{equation}
where $d_{br},\, d_{rd},\, d_{re},\, d_{bd}$, and $d_{be}$ are the Alice-RIS, RIS-Bob, RIS-Eve, Alice-Bob, and Alice-Eve distances, respectively, $\alpha_i,\, \beta_{di},\, \beta_{ei},\, \eta_{d}$, and $\eta_{e}$ are independent random variables following a Rayleigh distribution, and $\chi$ is the path loss exponent.

Then, the received SNR at Bob and Eve can be, respectively, written as follows
\begin{equation}\label{BobSNR}
\begin{aligned}
    \gamma_d = \dfrac{|({\bf{h}}_{rd}{\bf{Q}}{\bf{h}}_{r}+h_{d})x|^2}{N_o}
\end{aligned}
\end{equation}

\begin{equation}\label{EveSNR}
\begin{aligned}
    \gamma_e = \dfrac{|({\bf{h}}_{re}{\bf{Q}}{\bf{h}}_{r}+h_{e})x|^2}{N_o}.
\end{aligned}
\end{equation}

Our objective is to maximize a variant of the secrecy capacity in an alternating element-wise manner \cite{9014322}, i.e., we optimize the phase of each reflecting element while the remaining ones are kept constants, with an optimization coefficient $\alpha$ to control Eve's channel capacity. More precisely, the objective function is given by
\begin{equation}\label{maxSR}
\max_{\phi_k}C(\gamma_d,\gamma_e)=\max_{\phi_k} \{\log_2{(1+\gamma_d)}-\alpha\log_2{(1+\gamma_e)}\},
\end{equation}
Note that by varying the coefficient $\alpha$ between $0$ and $1$, we study the trade-off between employing the RIS to maximize the capacity of Bob's channel (usually when Eve's location is unknown), or maximizing the secrecy rate that takes into account both channels (case of $\alpha$ =1).We will show, via simulations, that $\alpha$ is a function of the considered scenario, i.e., the relative Bob and Eve's channel qualities.

Considering \eqref{BobSNR} and \eqref{EveSNR}, we can then formulate the following optimization problem 
\begin{equation}\label{objectiveFunction}
\max_{\phi_k}\max\left(\log_2\bigg( \dfrac{1+\dfrac{P_t}{N_o}|{\bf{h}}_{rd}{\bf{Q}}{\bf{h}}_{r}+h_{d}|^2}{(1+\dfrac{P_t}{N_o}|{\bf{h}}_{re}{\bf{Q}}{\bf{h}}_{r}+h_{e}|^2)^\alpha}\bigg),0\right)
\end{equation}
\begin{flalign*}
\qquad s.t.:\qquad &{\bf{Q}}=\text{diag}[e^{j\phi_1},\cdots,e^{j\phi_N}]&
\end{flalign*}
We denote the objective function as
\begin{equation}
g({\bf{Q}})\triangleq\dfrac{1+\dfrac{P_t}{N_o}|{\bf{h}}_{rd}{\bf{Q}}{\bf{h}}_{r}+h_{d}|^2}{(1+\dfrac{P_t}{N_o}|{\bf{h}}_{re}{\bf{Q}}{\bf{h}}_{r}+h_{e}|^2)^\alpha}\triangleq\frac{N({\bf{Q}})}{(D({\bf{Q}}))^\alpha},    
\end{equation}
and we try to find the roots of the derivative of $g({\bf{Q}})$ with respect to $\phi_k$ given the other phases $\phi_i, i \neq{k}$ as constants. The numerator at $\phi_k$, $N(\phi_k)$, can be written as follows
\begin{align}\label{N_phi}
    N(\phi_k) =1+\frac{P_t}{N_o}|h_d+h_{rk}e^{j\phi_{k}}h_{rdk}+\sum_{\substack{i=1\\i\neq{k}}}^{N}h_{ri}e^{j\phi_i}h_{rdi}|^{2},
\end{align}
using the identity $|A+B|^2=|A|^2+|B|^2+2Re\{A\bar{B}\}$ and defining
\begin{equation}
\sum_{\substack{i=1\\i\neq{k}}}^{N}\bar{h}_{ri}e^{-j\phi_i}\bar{h}_{rdi} \triangleq C_{Rd}+jC_{Id},\notag
\end{equation}
equation \eqref{N_phi}, with some manipulations, can be written as
\begin{equation}\label{N_phi_simple}
    N(\phi_k) = A_d + B_dcos(\phi_k)+C_dsin(\phi_k)
\end{equation}
where
\begin{equation*}
\begin{split}
A_d&\triangleq1+\dfrac{P_t}{N_o}\bigg(|h_d|^2+(|h_{rk}||h_{rdk}|)^2+|C_{Rd}+jC_{Id}|^2\\
&+2|h_d|(C_{Rd}cos(\tau_d)+C_{Id}sin(\tau_d))\bigg)
\end{split}
\end{equation*}
\begin{equation*}
\begin{split}
B_d&\triangleq2\dfrac{P_t}{N_o}|h_{rk}||h_{rdk}|\bigg(|h_d|cos(\theta_k-\tau_d+\psi_{dk})\qquad\quad\,\,\\
&+C_{Rd}cos(\theta_k+\psi_{dk})+C_{Id}sin(\theta_k+\psi_{dk})\bigg)
\end{split}
\end{equation*}
\begin{equation*}
\begin{split}
C_d&\triangleq2\dfrac{P_t}{N_0}|h_{rk}||h_{rdk}|\bigg(|h_d|sin(\theta_k-\tau_d+\psi_{dk})\qquad\quad\,\\
&+C_{Rd}sin(\theta_k+\psi_{dk})-C_{Id}cos(\theta_k+\psi_{dk})\bigg).
\end{split}
\end{equation*}

Similarly, the denominator $D(\phi_k)$ can be written as
\begin{equation}\label{D_phi_simple}
    D(\phi_k) = A_e + B_ecos(\phi_k)+C_esin(\phi_k),
\end{equation}
where
\begin{equation*}
\begin{split}
A_e&\triangleq1+\dfrac{P_t}{N_o}\bigg(|h_e|^2+(|h_{rk}||h_{rek}|)^2+|C_{Re}+jC_{Ie}|^2\\
&+2|h_e|(C_{Re}cos(\tau_e)+C_{Ie}sin(\tau_e))\bigg)
\end{split}
\end{equation*}
\begin{equation*}
\begin{split}
B_e&\triangleq2\dfrac{P_t}{N_o}|h_{rk}||h_{rek}|\bigg(|h_e|cos(\theta_k-\tau_e+\psi_{ek})\qquad\quad\,\,\\
&+C_{Re}cos(\theta_k+\psi_{ek})+C_{Ie}sin(\theta_k+\psi_{ek})\bigg)
\end{split}
\end{equation*}
\begin{equation*}
\begin{split}
C_e&\triangleq2\dfrac{P_t}{N_0}|h_{rk}||h_{rek}|\bigg(|h_e|sin(\theta_k-\tau_e+\psi_{ek})\qquad\quad\,
\\&+C_{Re}sin(\theta_k+\psi_{ek})-C_{Ie}cos(\theta_k+\psi_{ek})\bigg)
\end{split}
\end{equation*}
with
\begin{equation}
\sum_{\substack{i=1\\i\neq{k}}}^{N}\bar{h}_{ri}e^{-j\phi_i}\bar{h}_{rei} \triangleq C_{Re}+jC_{Ie},\notag
\end{equation}

Thus, the objective function at $\phi_k$, $g(\phi_k)$, is written as
\begin{equation}\label{Abbreviation}
g(\phi_k)=\frac{N(\phi_k)}{D(\phi_k)^\alpha}=\frac{A_e + B_ecos(\phi_k)+C_esin(\phi_k)}{(A_e + B_ecos(\phi_k)+C_esin(\phi_k))^\alpha},
\end{equation}
next, we equate the derivative of the objective function with zero as follows
\begin{equation}\label{gDerivative}
    \dfrac{dg(\phi_k)}{d\phi_k}=\dfrac{\dfrac{dN(\phi_k)}{d\phi_k}D(\phi_k)^\alpha-\dfrac{dD(\phi_k)^\alpha}{d\phi_k}N(\phi_k)}{(D(\phi_k)^\alpha)^2}=0
\end{equation}
which is equivalent to solving the following equation
\begin{equation}\label{function_to_optimize}
    D(\phi_k)\dfrac{dN(\phi_k)}{d\phi_k}=\alpha N(\phi_k)\dfrac{dD(\phi_k)}{d\phi_k}
\end{equation}

Finally, considering \eqref{N_phi_simple} and \eqref{D_phi_simple}, the left-hand-side and the right-hand-side of \eqref{function_to_optimize} can be written as follows 
\begin{equation}\label{LHS}
\begin{split}
    D(\phi_k)\dfrac{dN(\phi_k)}{d\phi_k}&=B_eC_dcos^2(\phi_k)-C_eB_dsin^2(\phi_k)\\
    &+(-B_eB_d+C_eC_d)cos(\phi_k)sin(\phi_k)\\
    &+A_e(C_dcos(\phi_k)-B_dsin(\phi_k))
\end{split}
\end{equation}
\begin{equation}\label{RHS}
\begin{split}
    \alpha N(\phi_k)\dfrac{dD(\phi_k)}{d\phi_k}&=\alpha\bigg(B_dC_ecos^2(\phi_k)-C_dB_esin^2(\phi_k)\\
    &+(-B_dB_e+C_dC_e)cos(\phi_k)sin(\phi_k)\\
    &+A_d(C_ecos(\phi_k)-B_esin(\phi_k))\bigg)
\end{split}
\end{equation}

Solving equation \eqref{function_to_optimize} is not straight forward, however, it involves only basic functions and can be easily solved numerically in a computationally tractable manner using MATLAB software with an acceptable accuracy.

\section{Results and Discussion}\label{sec4}
\begin{figure}[t]
\centering
\includegraphics[width=1\linewidth, trim={2.5in 1.8in 3in 2in},clip]{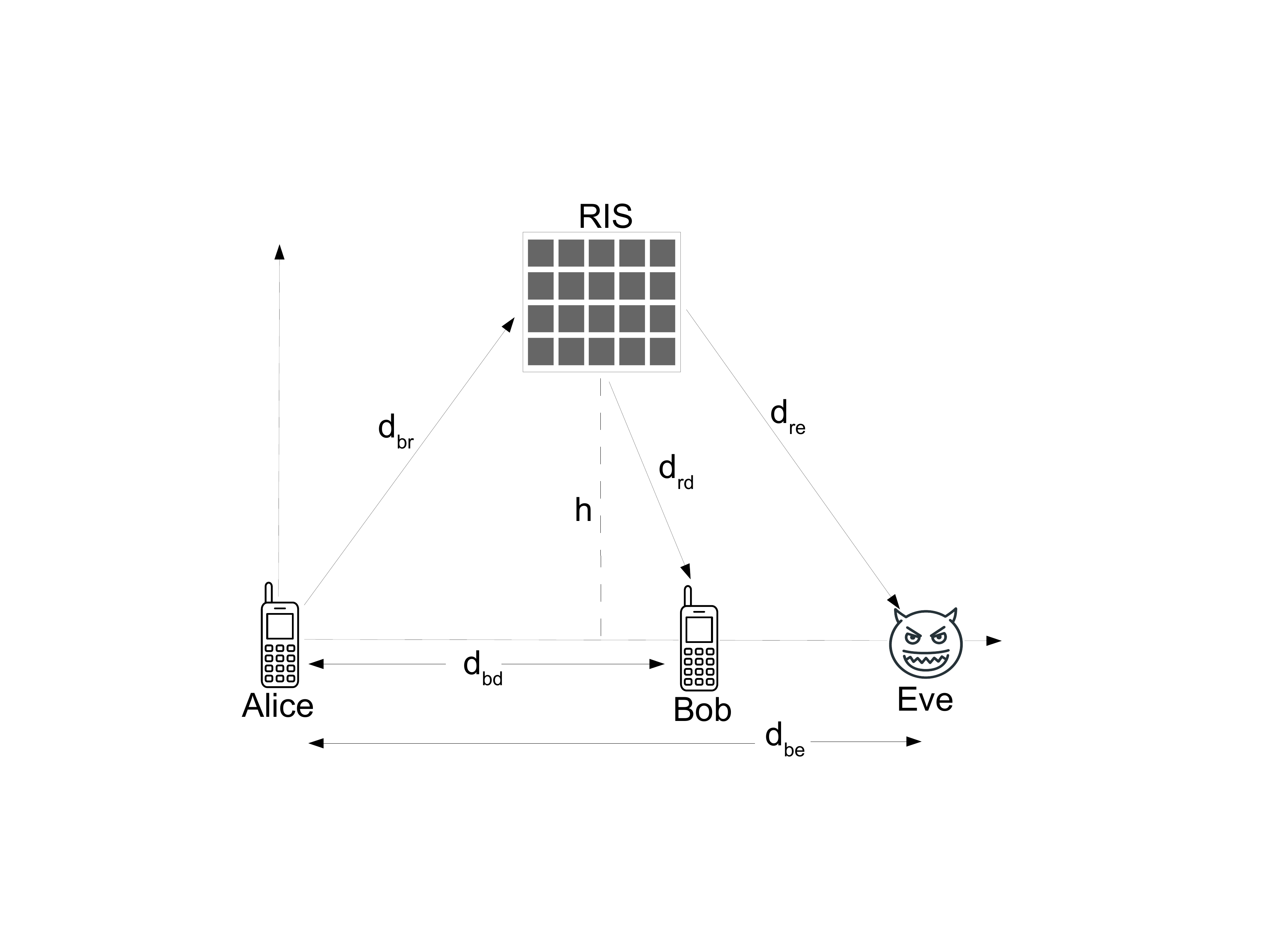} \\
\caption{Simulations setup}
\label{SimulationSetup}
\end{figure}

Simulation results are provided in this section to validate the methodology and to show the benefits of having an RIS for enhancing the secrecy of a communication system. Moreover, the effect of having direct links with Bob and Eve is shown. The path loss exponent, over all involved channels, is assumed fixed and is given as $\chi=3$,  and the noise power spectral density is set to $N_o = 10^{-10}$ Watts/Hz. Different setups are considered where Alice, Bob and Eve are assumed to be on a straight line in the vicinity of the RIS as shown in Fig.~\ref{SimulationSetup}. All simulation results are generated by averaging the secrecy capacity over $10^4$ channel realisations and $2$ optimization repetitions to alternatively improve the solution. The number of repetitions was chosen as we consider the case of having $N=2$ reflecting elements in the simulated setups.

\begin{figure}[!ht]
\centering
\includegraphics[width=0.9\linewidth, trim={0.2in 0in .4in 0.3in},clip]{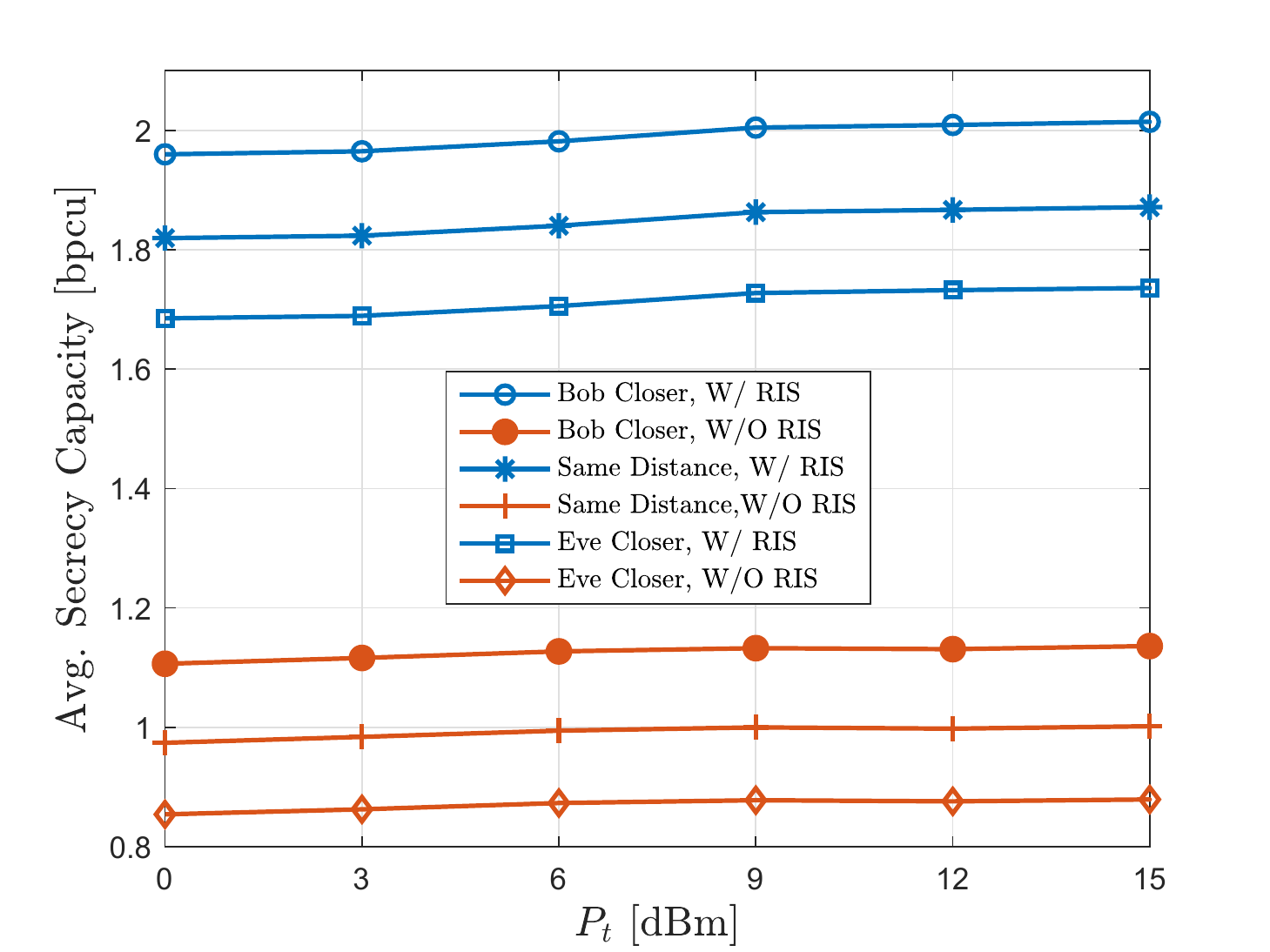} \\
\caption{Average secrecy capacity with $\alpha=1$ and $N=2$, with direct links and with/without RIS for different distance setups}
\label{setup1}
\end{figure}

In Fig. \ref{setup1}, assuming the existence of the direct links  Alice-Bob and Alice-Eve, we show the effect of having an RIS on the system secrecy capacity for different scenarios, where we set $\alpha=1$, i.e., the RIS has the CSI of both Bob's and Eve's channels, and $N=2$. Three different setups are considered, with either Alice-Bob or Alice-Eve average channel gain is better than the other or both have comparable average gain, which is mainly controlled by changing the distances Alice-Bob, $d_{bd}$ and Alice-Eve, $d_{be}$. It is clear that deploying an RIS improves the system secrecy capacity regardless of the setup. This is due to the assumption that the CSI of all involved channels is available at the RIS, and we optimise the RIS phase shifts to align the reflected signal with that of the direct path for Bob, while corrupting the signals at Eve. It is worth noting here that the existence of a relatively reliable direct link dominates the secrecy capacity which justifies the relatively small improvement in the secrecy capacity while increasing the transmit power.

\begin{figure}[!ht]
\centering
\includegraphics[width=0.9\linewidth, trim={0.2in 0in .4in 0.3in},clip]{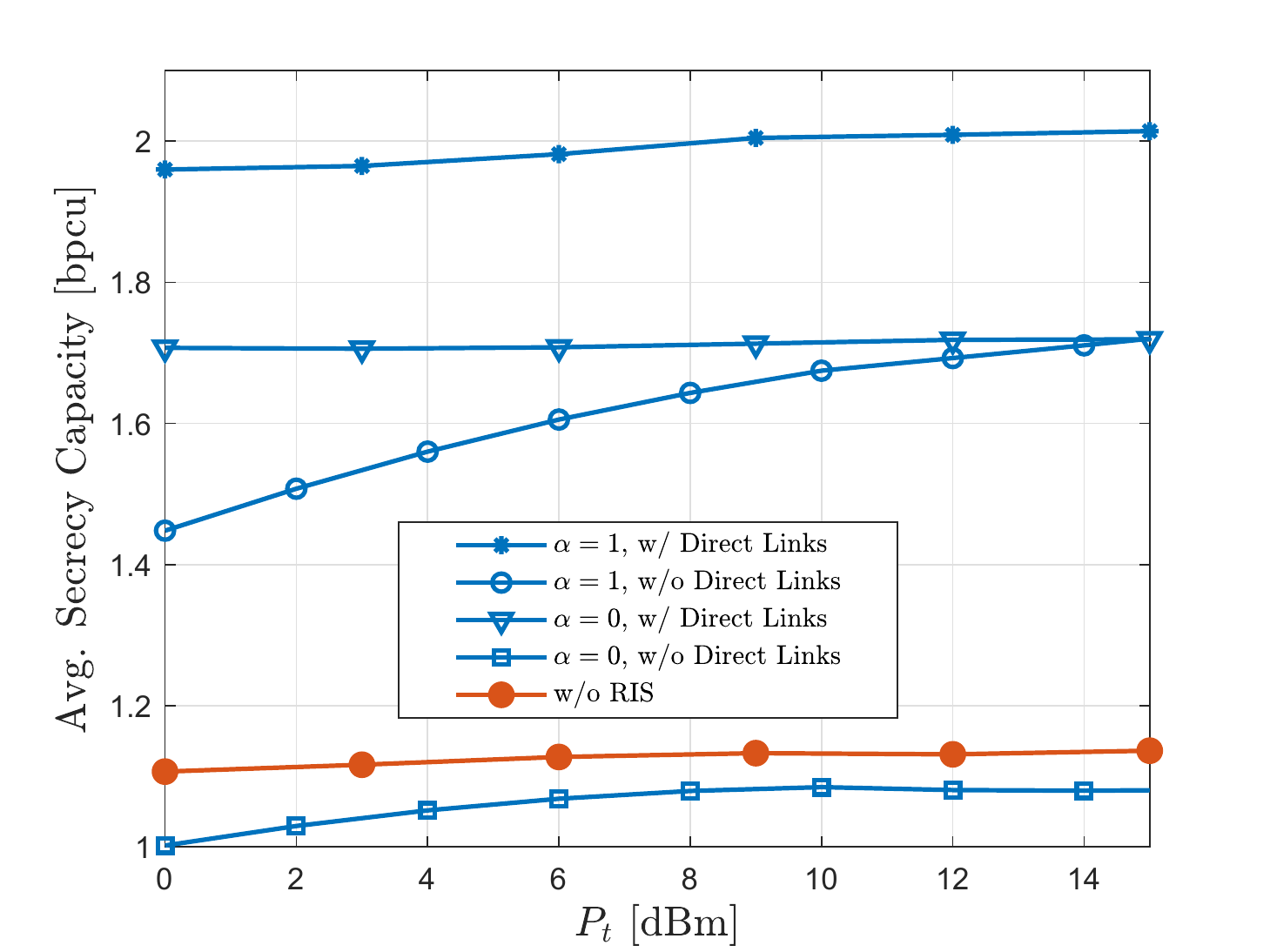} \\
\caption{Average secrecy capacity with $N=2$, $\alpha=0,1$ with and without the direct links}
\label{setup2}
\end{figure}

In Fig.~\ref{setup2}, we show the effect of having the direct links on the secrecy capacity and the practical case when the CSI of the RIS-Eve channel is unknown, i.e., $\alpha=0$. In this figure, Bob is assumed closer to Alice and hence having a better average channel gain. Overall, knowing the CSI of the RIS-Eve channel at the RIS improves the secrecy capacity as does having the direct links. One interesting notice here is that, with the RIS, and knowing the CSI of the RIS-Eve channel while the direct links are blocked, the system can achieve a comparable secrecy capacity to that when the CSI of RIS-Eve channel is unknown and the direct links do exist. Finally, without the knowledge of the RIS-Eve's CSI and with blocked direct links, having an RIS in the system does not improve the secrecy capacity and even worsen it depending on the relative RIS-Bob and RIS-Eve channel qualities as compared to that without an RIS.

\section{Conclusions and Future Work}\label{sec5}
The secrecy capacity of a SISO RIS-assisted cooperative network is investigated in this paper. Specifically, an optimization problem is formulated to maximize the secrecy capacity by optimizing the RIS-induced phase shifts taking into account the existence of the direct links from Alice to Bob and Eve. Furthermore, the effect of having the CSI of Eve's channel at the RIS is studied by introducing a weight $\alpha$ on Eve's channel capacity. Numerical simulations showed that the existence of reliable direct links dominates the system's secrecy capacity. However, it can be further improved by optimizing the RIS-induced phase shifts. Furthermore, the availability of the RIS-Eve CSI at the RIS boosts the RIS effectiveness. A more interesting finding is that, with the direct links are blocked and knowing the RIS-Eve CSI, the RIS-assisted system can achieve a comparable secrecy capacity to that with a dominant direct links and the RIS-Eve CSI is unavailable. However, in the numerical simulations we studied two extreme cases, $\alpha=0,1$, further investigations are required to find the optimal value of $\alpha$ in terms of the relative distances and the availability of Eve's CSI.

\section*{Acknowledgment}
The work of Prof. Hasna is supported by Qatar University research grant QUSM-CENG-2019-2. In addition, the work of Al-Kababji is supported by Qatar Foundation Graduate Sponsorship Research Award (GSRA6-2-0521-19034). Moreover, The HPC resources and services used in this work were provided by the Research Computing group in Texas A\&M University at Qatar. Research Computing is funded by the Qatar Foundation for Education, Science and Community Development (http://www.qf.org.qa). The statements made herein are solely the responsibility of the author[s].

\bibliographystyle{IEEEtran}
\bibliography{IEEEabrv.bib,RIS.bib}{}
\end{document}